# A Non-existence Proof of Quantum Phase Transition in Spin-Boson Model

Tao Liu[1], Zexian Cao[2]

[1]School of physics, Southwest University of Science & Technology, Mianyang 621000, China

[2]Institute of physics, Chinese Academy of Sciences, Beijing 100190, China

E-mail: liutao849@163.com; zxcao@iphy.ac.cn

**Abstract**

Quantum phase transition in the spin-boson model was claimed on the basis of various numerical studies, but not strictly proven. Here by using a unitary transformation to decompose the Hamiltonian into two branches of odd and even parity we obtained the necessary and sufficient condition for degeneracy to occur between states of opposite parity in the spin-boson model, and the analytical expression for such degenerate energies. It can be strictly proven that the ground state of spin-boson model with non-vanishing tunneling amplitude must have an energy lower than the lowest possible such degenerate energy, and have definite parity. Starting from the invariancy of the parity operator we show that finite expansion by numerical calculation induces the breaking of parity symmetry responsible for the phase transition. The critical dissipation parameter we obtained for parity-symmetry breaking, as a logarithmic function of summed diagonal matrix elements in the finite expansion for the bosonic part of the parity operator, can reproduce the phase diagram derived with quantum Monte Carlo method and logarithmically discretized numberical renormalization group approach. It reveals that the quantum phase transition in spin-boson model claimed by numerical procedures arises from symmetry breaking caused by finite expansion in practical calculation. The method we developed here may also be applicable to the discussion of quantum chaos and other similar problems.

# I. Introduction

Spin-boson model is a particular, two-level realization of the Caldeira-Leggett model which was devised three decades ago to describe quantum dissipation problem[1]. It can be used as a simple model to study the dynamics of a dissipative particle confined in a double-well potential, to represent a qubit coupled to an environment that can provoke decoherence, etc[2,3]. Moreover, it also represents a paradigm in the study of quantum phase transition. For a critical value of the coupling strength to the bath, a phase transition from a regime in which the particle is delocalized among two positions (or two spin states) to the region of localization has been claimed[3].

The paradigm Hamiltonian for the spin-boson model with zero external field is, setting $\hbar = 1$, given by

$$H_{sb} = -\frac{\Delta}{2}\sigma_x + \sum_k \omega_k a_k^+ a_k + \frac{1}{2}\sum_k \lambda_k (a_k^+ + a_k)\sigma_z, \quad (1)$$

where $\sigma_x$ and $\sigma_z$ are the standard Pauli matrices, $\Delta$ is the tunneling amplitude between the two levels of spin, $a_k^+$ and $a_k$ are the creation and annihilation operators of the bath modes with frequencies $\omega_k$, and $\lambda_k$ is the strength of coupling between spin and the $k$-th bath mode[4]. The effect of the harmonic oscillator environment is encoded in the spectral function $J(\omega) = \pi\sum_k \lambda_k^2 \delta(\omega - \omega_k)$ for $0 < \omega < \omega_c$, where $\omega_c$ is the cutoff energy. In the infrared limit, i.e., $\omega \to 0$, the power laws regarding the spectral function $J(\omega)$ are of particular importance. Considering the low-energy details of the spectrum, it has $J(\omega) = 2\pi\alpha\omega_c^{1-s}\omega^s$, here α is the dimensionless dissipation parameter related to $\lambda_k$ [4]. The exponent s characterizes the nature of different baths, with $s > 1$ for super-ohmic bath, $s = 1$ for ohmic bath, and $s < 1$ for sub-ohmic bath.

It was claimed that for ohmic and sub-ohmoc bathes, with the dissipation parameter increasing cross a critical value $\alpha_c$, quantum phase transition occurs from a delocalized, non-degenerate ground state with zero magnetization to a localized, twofold-degenerate ground state with non-zero magnetization[5-7]. So far the quantum phase transition has been revealed on the basis of numerical studies which employed the numerical renormalization group (NRG) technique[5,6,8-14], the quantum Monte Carlo (QMC) method[7], an approach that combines polynomial expansion of spectral function with the sparse grid concept[15], the density matrix renormalization group method[16], and the variational matrix-product-state approach[17], etc. Analytical studies of spin-boson model are also available, but seem not to be concerned with the existence of the transition. By means of a perturbative approach based on a unitary transformation[18,19] the decoherence of the two-state system coupled with a sub-Ohmic bath was investigated, revealing that when the system undergoes a transition (crossover) from the delocalized state to the localized state, the time evolution of the two-level system changes from coherent to decoherent dynamics. A latest work applying the Silbey-Harris variational polaron ansatz describes the quantum phase transition with mean-field exponent for the subohmic case ( $0 < s < 0.5$ ), where variation is done with the magnetization *a priori* set as a non-zero constant[20].

It is safe to say that so far all the clues or evidences pointing to the occurrence of ground state degeneracy with increasing coupling strength in the spin-boson model come from numerical results, and the properties of the associated quantum phase transition are extracted by numerical approaches, by which various approximation schemes have to be adopted. Even the analytical approaches also have to deploy finite expansion, for instance, taking the zero-order expansion of coherent state as wavefunction to solve the problem[20]. There is no strict existence proof for ground state degeneracy involing opposite parity in the spin-boson model. The work of Spohn on ground state(s) of the spin-boson Hamiltonian is often taken as the existence proof for quantum phase transition in this model, but we found it not impeccable as there a new algrebraic space was introduced to resolve the divergence problem of averaged phonon number, and the solution for the stationary Schrödinger equation is not in the complete set defined by the Hamiltonian, which is certainly unacceptable [21]. A question of fundamental importance, yet seemingly not seriously addressed, may be raised: is there ground state degeneracy involving opposite parity, which is essential for quantum phase transition in spin-boson model, as a true intrinsic property of the system or being simply a numerical artifact? Due to the enormous degrees of freedom for such an open quantum system which involves multiple-mode harmonic oscillators (the degrees of freedom for the bosonic field is given by $(N+1)^M$, where M is the number of modes, N the number of bosons for each mode), the proof of this problem by numerical approaches is formidable. On the other hand, the ground state degeneracy problem will in principle lead to a characteristic equation of a degree much larger than quintic, which cannot be solved by an algebraic formula that involves only simple operations. It seems we have come to a cul-de-sac.

This dilemma may be circumvented the other way. If we are able to obtain the analytical expression of the degenerate energies referring to states of opposite parity for the spin-boson model, and compare the ground state energy with the lowest possible such degenerate energy, then a criterion can be established to help answer the above question. Fortunately, since the Hamiltonian of the spin-boson model for the case of zero external field has parity symmetry, the complete set of all the possibly degenerate energies for both odd and even branches of the spectrum can be analytically obtained. The parity of the system is related to its state, and a state of definite parity has always $<\sigma_z>=0$.

In this article, we give a strict non-existence proof for the quantum phase transition in spin-boson model. By using a unitary transformation the original Hamiltonian is decomposed into two branches of distinct parity, and subsequently the necessary and sufficient condition for the occurrence of degeneracy between states of opposite parity is derived. An auxiliary form of Hamiltonian is constructed as the direct-sum of the Hamiltonians of distinct parity, which gives the analytical expression of the degenerate energies referring to states of opposite parity. By applying the Rayleigh quotient of matrix algebra, the ground state energy is shown to be smaller than the lowest possible such degenerate energy, thus complete the non-existence proof for the ground state degeneracy in spin-boson model. Moreover, starting from the parity invariance, by which we mean that $P^2 = 1$ holds in all representations [22] of the spin-boson model, we can prove that the finite expansion of our parity operator manifests a logarithmic parity-breaking critical point, that when the dissipation parameter α increases over this critical point, symmetry breaking occurs. The quantum phase critical points obtained in QMC and NRG are quite consistent, regarding to its behavior following the choice of cutoff for the number of bosonic modes or the number of bosons on individual bosonic modes, with our result, confirming that the so-called quantum phase transition for spin-boson model is the consequence of symmetry-breaking induced by finite expansion, rather than intrinsic property of the system.

**II. Non-existence proof**

For the Hamiltonian in (1) there exists the parity operator $P_{sb} = \sigma_x e^{i\pi \sum_k a_k^+ a_k}$, which commutes with the Hamiltonian, i.e., $[P_{sb}, H_{sb}] = 0$. With the unitary transformation $U = \frac{1}{\sqrt{2}} \begin{pmatrix} 1 & e^{-i\pi \sum_k a_k^+ a_k} \\ -1 & e^{i\pi \sum_k a_k^+ a_k} \end{pmatrix}$ it has $P = U P_{sb} U^{-1} = (e^{i\pi \sum_k a_k^+ a_k})^2 \sigma_z = \sigma_z$, and

$$H = U H_{sb} U^{-1} = \begin{pmatrix} H^+ & 0 \\ 0 & H^- \end{pmatrix}, \text{ with}$$

$$H^+ = H_0 - H_\Delta \tag{2a}$$

$$H^- = H_0 + H_\Delta \tag{2b}$$

where $H_0 = \sum_k \omega_k [A_k^+ A_k - q_k^2]$, $H_\Delta = \Delta e^{i\pi \sum_k a_k^+ a_k} / 2$, in which $A_k^+ = a_k^+ + q_k$, $A_k = a_k + q_k$, and $q_k = \lambda_k / 2\omega_k$.

The unitary transformation $U$ turns the original Hamiltonian into two branches, $H^+$ of even parity and $H^-$ of odd parity, of which the corresponding energy spectrum can be denoted as $E^\pm$. Suppose $|\psi\rangle = \begin{pmatrix} |\varphi^+\rangle \\ |\varphi^-\rangle \end{pmatrix}$, obviously it has $P \begin{pmatrix} |\phi^+\rangle \\ 0 \end{pmatrix} = \begin{pmatrix} |\phi^+\rangle \\ 0 \end{pmatrix}$,

$$P\begin{pmatrix} 0 \\ |\phi^-\rangle \end{pmatrix} = -\begin{pmatrix} |0\rangle \\ |\phi^-\rangle \end{pmatrix}, \quad \text{and} \quad H^\pm|\varphi^\pm\rangle = E^\pm|\varphi^\pm\rangle .$$ Thus, $H^+$ and $H^-$ are mutually independent, each having its own sub-space of definite parity. Without loss of generality, it can be assumed that the stationary wavefunctions for the Schrödinger equation corresponding to $H^+$ and $H^-$ are

$$|\varphi^+\rangle = \sum_{\{n\}} c^+_{\{n\}} |\{n\}\rangle_A , \qquad (3a)$$

$$|\varphi^-\rangle = \sum_{\{n\}} c^-_{\{n\}} |\{n\}\rangle_A , \qquad (3b)$$

respectively. Here $\{n\} = n_1, \cdots, n_M$ is the combination of numbers of boson on M modes. $|\{n\}\rangle_A$ is the displaced Fock states $|\{n\}\rangle_A = \prod_{k=1}^M |n_k\rangle_{A_k}$, in which

$$|n_k\rangle_{A_k} = \frac{(a_k^+ + q_k)^{n_k}}{\sqrt{n_k!}} e^{-q_k a_k^+ - q_k^2/2} |0\rangle , \qquad \text{satisfying} \qquad {}_A\langle\{m\}|\{n\}\rangle_A = \delta_{\{m\},\{n\}} \quad \text{and}$$

$\sum_{\{n\}} |\{n\}\rangle_A {}_A\langle\{n\}| = 1$.

Putting the Hamiltonians in (2a-2b) and the corresponding wavefunctions in (3a-3b) into the Schrödinger equation, one obtains

$$\sum_k \omega_k [A_k^+ A_k - q_k^2] |\varphi^+\rangle - \frac{\Delta}{2} e^{i\pi \sum_k a_k^+ a_k} |\varphi^+\rangle = E^+ |\varphi^+\rangle , \qquad (4a)$$

$$\sum_k \omega_k [A_k^+ A_k - q_k^2] |\varphi^-\rangle + \frac{\Delta}{2} e^{i\pi \sum_k a_k^+ a_k} |\varphi^-\rangle = E^- |\varphi^-\rangle . \qquad (4b)$$

Multiplying eqs.(4a-4b) from left side with $\langle\phi^-|$ and $\langle\phi^+|$, respectively, it leads to

$$\langle\varphi^-|H_0|\varphi^+\rangle - \langle\varphi^-|H_\Delta|\varphi^+\rangle = E^+ \langle\varphi^-|\varphi^+\rangle , \qquad (5a)$$

$$\langle\varphi^+|H_0|\varphi^-\rangle + \langle\varphi^+|H_\Delta|\varphi^-\rangle = E^- \langle\varphi^+|\varphi^-\rangle . \qquad (5b)$$

Since from eqs.(5a-5b) it has $E^- - E^+ = 2\langle\phi^+|H_\Delta|\phi^-\rangle / \langle\phi^+|\phi^-\rangle$ (for justification, see appendix), clearly

$$\langle\varphi^+|H_\Delta|\varphi^-\rangle = 0 \qquad (6)$$

is the necessary and sufficient condition for the occurrence of energy degeneracy with regard to $H^+$ and $H^-$ which reside independently in the odd and even subspaces of a complete set for the spin-boson model.

By constructing the direct sum for the Hamiltonians $H^+$ and $H^-$, all the possible degenerate energies for states of opposite parity, i.e., energies to appear in the energy spectra of both $H^+$ and $H^-$, can be obtained. Toward this end, let's denote

$$H_1 = H^+ \otimes I + I \otimes H^- \tag{7a}$$

$$H_2 = H^+ \otimes I - I \otimes H^- . \tag{7b}$$

Since $[H_1, H_2] = 0$, thus $H_1$ and $H_2$ have common eigenstates which, without any loss of generality, can be given in the form

$$|\varphi\rangle = |\varphi^+\rangle \otimes |\varphi^-\rangle = \sum_{\{i\},\{j\}} f_{\{i,j\}} |\{i\}\rangle_A |\{j\}\rangle_A . \tag{8}$$

Then the Hamiltonians $H_1$ and $H_2$ in (7a-7b) satisfy the equations

$$H_1 |\varphi\rangle = E_1 |\varphi\rangle , \tag{9a}$$

$$H_2 |\varphi\rangle = E_2 |\varphi\rangle , \tag{9b}$$

where $E_1 = E^+ \oplus E^-$, $E_2 = E^+ \oplus (-E^-)$. This is to say that the eigenvalues of $H_1$ are the sum of those for $H^+$ and $H^-$, and the eigenvalues of $H_2$ are the sum of those for $H^+$ and $-H^-$. Adding eq.(9a) and eq.(9b), one obtains

$$[H_0 \otimes I - (E_1 + E_2)/2] |\varphi\rangle = (H_\Delta \otimes I) |\varphi\rangle . \tag{10}$$

For the equation

$$(H_\Delta \otimes I) |\varphi\rangle = 0 \tag{11}$$

derived from a vanishing right-hand side of eq.(10) to have definite solution, the necessary and sufficient condition is $\det(H_\Delta \otimes I) = 0$.

It can be easily proven that eq.(11) satisfies (6) which is the necessary and sufficient condition for energy egeneracy to occur between $H^+$ and $H^-$. This is because, by multiplying (11) from left side with ${}_A\langle n|{}_A\langle m|$, it thus formally has

$$\Delta \sum_{\{i\}} f_{\{i,n\}} D_{\{m,i\}} = 0 , \tag{12}$$

here n, m can be infinite; for practical computation, $\{m\},\{n\}=\{0\},\ldots\{N_{max}\}$, $N_{max}$ is a cutoff value, consequently it must have $\Delta \sum_{\{i,n\}} f_{\{i,n\}} D_{\{n,i\}} = 0$. And since

$$\Delta\langle\varphi^+|e^{i\pi\sum_k a_k^+ a_k}|\varphi^-\rangle = \Delta\sum_{\{m\},\{n\}} c^+_{\{m\}} c^-_{\{n\}} D_{\{m,n\}} \quad, \quad \text{setting} \quad f_{\{m,n\}} = c^+_{\{m\}} c^-_{\{n\}} \quad, \quad \text{it has}$$

$$\Delta\sum_{\{m\},\{j\}} f_{\{m,j\}} D_{\{m,j\}} = \Delta\langle\varphi^+|e^{i\pi\sum_k a_k^+ a_k}|\varphi^-\rangle = 0.$$

Therefore, for a non-vanishing solution $|\varphi\rangle$ that satisfies (11), it has $E^+ = E^-$. Denoting such $E^+ = E^-$ as $E^{eo}$, then $E_1 = 2E^{eo}$, $E_2 = 0$, and eq.(10) turns into

$$[H_0 \otimes I - E^{eo}]|\varphi\rangle = 0. \tag{13}$$

Noticing that $H_0 \otimes I$ as a matrix is diagonal in the space of $|\{i\}\rangle_A |\{j\}\rangle_A$, thus all the possible $E^{eo}$ values form the complete set of eigenvalues for $H_0$, and the analytical expression for the individual eigenvalues is

$$E^{eo}_{\{m\}} = \left(\sum_k \omega_k m_k\right)_{\{m\}} - \sum_k \omega_k q_k^2, \quad m = 0,1...\infty. \tag{14}$$

We see that under the necessary and sufficient condition for energy degeneracy to occur between $H^+$ and $H^-$, all the possible degenerate energies involving opposite parity are contained in the complete set of eigenvalues for $H_0$, independent from the tunneling amplitude $\Delta$. Or, the concrete degenerate energy for $H^+$ and $H^-$ with respect to different $\Delta$ must be contained in the complete set of eigenvalues for $H_0$, which in principle can be determined from eq.(14) together with eq.(4a) or eq.(4b).

The complete set of degenerate energy involving opposite parity given by eq.(14) is of essential importance for the judgment whether such degeneracy occurs to the ground state with increasing dissipation parameter α：according to eq.(14), the lowest possible such degenerate energy is $E^{eo}_{min} = E^{eo}_{\{0\}}$. Below we can prove that for $\Delta \neq 0$, the minimum energy for the system to have a definite parity satisfies $\text{Inf}(E^+_{min}, E^-_{min}) < E^{(eo)}_{min}$, here $E^+_{min}$ and $E^-_{min}$ are the respective minimum for the eigenvalues of $H^+$ and $H^-$. This is to say that the smaller one of $E^+_{min}$ and $E^-_{min}$ must fall outside the complete set of degenerate energy for $H_0$.

For convenience, let's denote $\varepsilon_i^{(\pm)}$ as the i-th eigenvalue of $H^+ + H^-$. For the hermitian matrices corresponding to $H^+$ and $H^-$, from the Rayleigh quotient of matrix algebra it has $E_{min}^+ + E_{min}^- \le \varepsilon_i^{(\pm)}$. Denote the minimum of $\varepsilon_i^{(\pm)}$ as $\varepsilon_{min}^{(\pm)}$, thus it has

$$E_{min}^+ + E_{min}^- \le \varepsilon_{min}^{(\pm)} = 2E_{min}^{eo}. \tag{15}$$

Some remarks are appropriate at this place.

1) $E_{min}^{eo}$ is the lowest possible degenerate energy for $H^+$ and $H^-$. Only when $E_{min}^+ = E_{min}^-$ the equality in (15) holds;

2) the necessary and sufficient condition for $E_{min}^+ = E_{min}^- = E_{min}^{eo}$ is $\Delta=0$.

Thus for the ground state energy it has

$$E_{gs} = \text{Inf}(E_{min}^+, E_{min}^-)\begin{cases} < E_{min}^{eo} & \text{if } \Delta \ne 0 \\ = E_{min}^{eo} & \text{if } \Delta = 0 \end{cases}, \tag{16}$$

and the following conclusion can be drawn: When the tunneling strength $\Delta=0$, the system has parity symmetry, $E_{min}^+ = E_{min}^-$, the ground state shows degeneracy involving opposite parity, $E_{gs} = E_{min}^{eo}$; but when $\Delta \ne 0$, $E_{min}^+ \ne E_{min}^-$ and $E_{gs} = \text{Inf}(E_{min}^+, E_{min}^-) < E_{min}^{eo}$. This is to say that so long as $\Delta \ne 0$, despite of the strength of dissipation, the tunneling can always lift the degeneracy of ground state for $\Delta=0$, thus to assume a definite parity. However, once the ground state assumes a definite parity, it has a vanishing magnetization ($<\sigma_z>=0$). Therefore the spontaneous occurrence of magnetization at ground state, i.e., the transition from the delocalized phase $<\sigma_z>=0$ to the localized phase $<\sigma_z>\ne 0$, which is the character of the quantum phase transition for spin-boson model, cannot occur.

The procedure we used here, i.e., by applying the unitary transformation to find and then compare the energy levels of individual parities, provides a method for judging the ground state degeneracy involving opposite parity for systems like spin-boson model. It is demonstrated for the single-mode case, but can be easily generalized to the multi-mode cases. This procedure may also be applicable to the discussion of quantum chaos and other relevant problems.

**III. Numberical demonstration**

The necessary and sufficient condition for degeneracy to occur between $H^+$ and $H^-$, which nest in the subspace of odd or even parity, respectively, is independent from the number of modes or the spectral distribution concerned. The complete set of degenerate energies involving opposite parity given in (14) is derived for the case of a single mode, but the procedure above is also applicable to the cases of arbitrary discrete or continuous phonon energy spectra. Particularly, the lowest possible degenerate energy between $H^+$ and $H^-$ in different cases can be given in a simple analytical form

$$E_{\min}^{eo} = \begin{cases} -\lambda^2/4\omega \\ -\sum_{k=1}^{N} \lambda_k^2/4\omega_k \\ -\alpha\omega_c/2s \end{cases} . \qquad (17)$$

Based on the aforementioned discussion it can be concluded that at $\Delta \neq 0$ the ground state of spin-boson model has a definite parity. However, numerical investigations in the past three decades based on various different approaches all pointed to the existence of a quantum phase transition from non-degenerate ground state of vanishing magnetization to the degenerate ground state of finite magnetization. The origin of this controversy now can be addressed.

First, with a complete orthonormal set of the wavefunctions for spin-boson model, the bosonic part of the parity operator $P = UP_{sb}U^+ = (e^{i\pi\sum_k a_k^+ a_k})^2 \sigma_z = \sigma_z$, that's $(e^{i\pi\sum_k a_k^+ a_k})^2$, is a unity matrix in matrix representation. It is independent from the dissipation parameter $\alpha$ or the exponent $s$ characterizing the spectral distribution. When expanded in the complete orthonormal set $\{|\{n\}\rangle_A\}$, this is

$$(e^{i\pi \sum_k a_k^+ a_k})^2 = \left( \sum_{\{n=0\}}^{\{\infty\}} D_{\{m,n\}} D_{\{n,m\}} |\{m\}\rangle_{AA} \langle m| \right) \quad , \tag{18}$$

$$= e^{-4\sum_k q_k^2} \left( \sum_{\{n=0\}}^{\{\infty\}} L_{\{m,n\}} L_{\{n,m\}} |\{m\}\rangle_{AA} \langle m| \right) = I$$

where $L_{\{m,n\}} = \prod_k \sum_{j=0}^{\text{Inf}[m_k,n_k]} \dfrac{(-1)^j \sqrt{m_k! n_k!} (2q_k)^{m_k+n_k-2j}}{(m_k - j)!(n_k - j)! j!}$. This is to say that in the complete set, $\{|\{n\}\rangle_A, \ n=0,...\infty\}$, the invariance of this bosonic part is preserved [22]. However, this cannot be guaranteed, to the precision preset for calculation, in the finite subspace $\{|\{n\}\rangle_A, \ n=0,...N_{tr}\}$.

In order to clarify this problem, let's check the diagonal matrix elements in eq.(18) $O_{\{m,m\}}^{\infty} \equiv \sum_{\{n=0\}}^{\{\infty\}} L_{\{m,n\}} L_{\{n,m\}}$ and its truncation $O_{\{m,m\}}^{N_{tr}} \equiv \sum_{\{n=0\}}^{\{N_{tr}\}} L_{\{m,n\}} L_{\{n,m\}}$. Obviously it has $O_{\{m,m\}}^{\infty} \geq O_{\{m,m\}}^{N_{tr}}$. The invariance of eq.(18) requires $e^{-4\sum_k q_k^2} O_{\{m,m\}}^{\infty} = 1$, i.e., $O_{\{m,m\}}^{\infty} e^{-2\alpha\beta} = 1$, here $2\alpha\beta = 4\sum_k q_k^2$ (for spectrum $J(\omega) = 2\pi\alpha\omega_c^{1-s}\omega^s$, $\beta = \omega_c^{1-s} \int_0^{\omega_c} \omega^{s-2} d\omega$ ). But, for any practical caclucation the truncation of this summation at a chosen $N_{tr}$ is necessitated, thus it has always $O_{\{m,m\}}^{N_{tr}} e^{-2\alpha\beta} \leq 1$. Denoting $\alpha_c = \ln(O_{\{m,m\}}^{N_{tr}})/2\beta$, thus it has

$$\alpha \geq \alpha_c \equiv \ln(O_{\{m,m\}}^{N_{tr}})/2\beta. \tag{19}$$

For a given dissipation parameter α and a sufficiently large $N_{tr}$, the condition $\alpha = \alpha_c$ can be always satisfied. Under this circumstance the invariance of the parity operator will be preserved. But for any given truncation at $N_{tr}$, when $\alpha > \alpha_c$, the invariance $(e^{i\pi \sum_k a_k^+ a_k})^2 = I$ in the subspace of the orthonormal complete set of wavefunctions is spoiled. The parity symmetry of the Hamiltonian suffers a breaking at $\alpha > \alpha_c$, but originating in the finite expansion, and $\alpha_c$ is the critical dissipation parameter governing the occurrence of symmetry-breaking of the system in this context. Briefly speaking, so long as $\alpha > \alpha_c$ holds, parity symmetry breaking occurs, but it is obviously brought about by the truncation procedure which introduces in the course of calculation a parity breaking to the ground state. Ignoring the origin of this parity-breaking mechanism, the spontaneous symmetry breaking in the magnetization $<\sigma_z>$ with increasing dissipation parameter $\alpha$, which was recognized in the calculation result, was then categorized as a quantum phase transition. To confirm this view, we compare our analytical result in (19) with those, for instance, derived by applying the NRG and QMC approaches (see Fig.1).

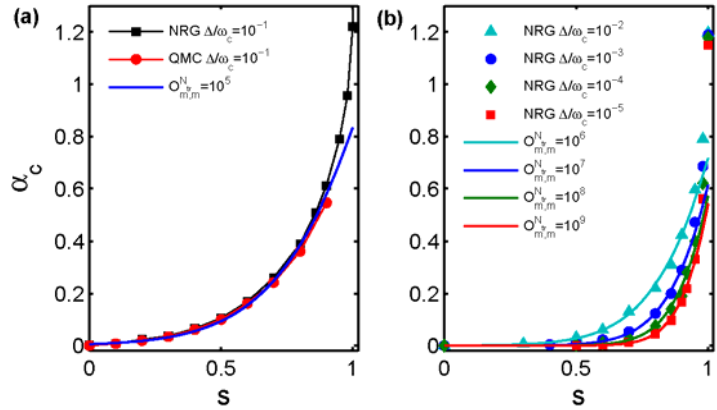

Fig. 1. The parity symmetry-breaking critical point $\alpha_c$ calculated from formula (19) for finite truncations and derived in the QMC and logarithmically discretized NRG approaches. (a) QMC and NRG data taken from Ref.[7]; (b) NRG results provided by Ning Hua Tong.

From Fig.1 we see that the phase diagram of the spin-boson model, i.e., the $\alpha_c \sim s$ curve, based on numerical data obtained by using NRG and QMC approaches can be well reproduced by the expression in eq.(19). This consistency is in no sense a coincidence. It clearly shows that the parity symmetry-breaking of the system in those works was caused by finite expansion introduced in practical calculation. The quantum phase transition claimed on the basis of numerical results should orginate in the 'backdoor' in the numerical procedure (see appendix).

**IV. Conclusion**

In summary, by transforming the Hamiltonian into two branches of distinct parity, followed by construction of auxiliary Hamiltonians in a direct-sum form and application of Rayleigh quotient, the necessary and sufficient condition for the spin-boson model to have degenerate energies involving opposite parity is formulated, and all such degenerate energies are analytically given. For all non-vanishing tunneling amplitude, the ground state of definite parity must have an energy less than the least possible such degenerate energy. This is to say that in principle there is no spontaneous quantum phase transition following increasing dissipation parameter in the spin-boson model. The quantum phase transition in spin-boson model concluded from numerical approaches originates in the parity symmetry breaking induced by finite expansion which is necessitated in the practical computation. The procedure developed here to accomplish the non-existence proof can also be applied to the discussion of other similar problems.

**Appendix**

**1. Proof of** $E^- - E^+ = \Delta \langle \varphi^+ | e^{i\pi \sum_k a_k^+ a_k} | \varphi^- \rangle$

First of all, the matrix form of the Hamiltonians $H^+$ and $H^-$ in the space of operator $A_k$ is a real symmetric matrix, this is because

$$\sum_k \omega_k A_k^+ A_k = \sum_{\{n=0\}}^{\{\infty\}} |\{n\}\rangle_{AA}\langle\{n\}| \sum_k \omega_k A_k^+ A_k = \sum_{\{n=0\}}^{\{\infty\}} \left(\sum_k \omega_k n_k\right)_{\{n\}} |\{n\}\rangle_{AA}\langle\{n\}|, \quad (A1)$$

$$e^{i\pi \sum_k a_k^+ a_k} = \sum_{\{m=0\},\{n=0\}}^{\{\infty\},\{\infty\}} D_{\{m,n\}} |\{m\}\rangle_{AA}\langle\{n\}|, \quad (A2)$$

and

$$H^\pm = \sum_{\{n=0\}}^{\{\infty\}} \left(\sum_k \omega_k n_k\right)_{\{n\}} |\{n\}\rangle_{AA}\langle\{n\}| - \sum_k \omega_k q_k^2 \mp \frac{\Delta}{2} \sum_{\{m=0\},\{n=0\}}^{\{\infty\},\{\infty\}} D_{\{m,n\}} |\{m\}\rangle_{AA}\langle\{n\}|. \quad (A3)$$

Noticing that $\left(\sum_k \omega_k n_k\right)_{\{n\}}$, $\sum_k \omega_k q_k^2$, $D_{\{m,n\}} = D_{\{n,m\}}$ are all real numbers, thus $H^\pm$ are real symmetric matrices. Here $D_{\{n,m\}}$ satisfies the relation $\sum_{\{j=0\}}^{\{\infty\}} D_{\{m,j\}} D_{\{j,n\}} \equiv \delta_{m,n}$, since

$$e^{i\pi \sum_k a_k^+ a_k} e^{i\pi \sum_k a_k^+ a_k}$$
$$= \sum_{\{m,n=0\}}^{\{\infty\}} D_{\{m,j\}} |\{m\}\rangle_{AA}\langle\{j\}| \sum_{\{k,n=0\}}^{\{\infty\}} D_{\{k,n\}} |\{k\}\rangle_{AA}\langle\{n\}|$$
$$= \sum_{\{m,n=0\}}^{\{\infty\}} \left( \sum_{\{j=0\}}^{\{\infty\}} D_{\{m,j\}} D_{\{j,n\}} \right) |\{m\}\rangle_{AA}\langle\{n\}| \qquad (A4)$$
$$= I$$

Accordingly, the eigenvector $|\varphi^\pm\rangle$ under the basis of $\{|\{n\}\rangle_A\}$ must be real, hence it must have

$$\langle\varphi^-|\varphi^+\rangle \equiv \langle\varphi^+|\varphi^-\rangle, \qquad (A5)$$

$$\langle\varphi^-|\sum_k \omega_k [A_k^+ A_k - q_k^2]|\varphi^+\rangle \equiv \langle\varphi^+|\sum_k \omega_k [A_k^+ A_k - q_k^2]|\varphi^-\rangle. \qquad (A6)$$

Based on these relations, subtracting eq.(4b) from eq.(4a) leads to

$$E^- - E^+ = \Delta\langle\varphi^+|e^{i\pi\sum_k a_k^+ a_k}|\varphi^-\rangle / \langle\varphi^+|\varphi^-\rangle.$$

Q.E.D.

## 2. Solution of characteristic equation in a subspace of the complete set

Not any arbitrary complete set can guarantee that the characteristic equation derived from the Schrödinger equation has single unique solution in the subspace of that complete set.

Taking the spin-boson model as an example,

$$H_{sb} = -\frac{\Delta}{2}\sigma_x + \sum_k \omega_k a_k^+ a_k + \frac{1}{2}\sum_k \lambda_k (a_k^+ + a_k)\sigma_z \qquad (B1)$$

Under the Fock states the general form of wavefunction is

$$|\psi_{fs}\rangle = \sum_{\{n=0\}}^{\{N_{tr}\}} \begin{pmatrix} c_{\{n\}}|\{n\}\rangle \\ d_{\{n\}}|\{n\}\rangle \end{pmatrix} \qquad (B2)$$

the characteristic equation for the stationary Schrödinger equation is

$$-\frac{\Delta}{2}d_{\{m\}} + \left(\sum_k \omega_k m_k\right)_{\{m\}} c_{\{m\}} + \frac{1}{2}[\left(\sum_k \lambda_k \sqrt{m_k+1}\right)_{\{m\}} c_{\{m+1\}} + \left(\sum_k \lambda_k \sqrt{m_k}\right)_{\{m\}} c_{\{m-1\}}] = Ec_{\{m\}}, \quad m=0\cdots N_{tr} \quad (B3)$$

$$-\frac{\Delta}{2}c_{\{m\}} + \left(\sum_k \omega_k m_k\right)_{\{m\}} d_{\{m\}} - \frac{1}{2}[\left(\sum_k \lambda_k \sqrt{m_k+1}\right)_{\{m\}} d_{\{m+1\}} + \left(\sum_k \lambda_k \sqrt{m_k}\right)_{\{m\}} d_{\{m-1\}}] = Ed_{\{m\}}$$

Eq.(B3) is not closed in the subspace of the complete set. It is solvable unless the terms involving $c_{\{m+1\}}$ and $d_{\{m+1\}}$ are ignored. The ratio of the number of undetermined coefficients discarded over the number of independent equations is $R = M/(N_{tr}+1)$, where M is the number of the total bosonic modes.

For a system of linear equations to have single unique solution, the number of unkonws, here the undetermined coefficients, and the number of independent equations must be equal. This condition can be satisfied by increasing $N_{tr}$ such that

$$R = \frac{M}{N_{tr}+1} \to 0 \qquad \text{( B4 )}$$

But for a sufficiently large M, it is impractical to choose a $N_{tr}$ such that it meets the condition (B4). The non-closed characteristic equation has no single unique solution.

If a complete set of states which cannot warrant the closure of the characteristic equation in the subspace were chosen for the study of spontaneous symmetry breaking in the ground state of a multiple-mode spin-boson model, then it could not be guaranteed that the results obtained via expansion in this finite subspace are irrelevant with the choice of the subspace.

This is to say that before we start the study of ground state of the spin-boson model, the 'backdoor' left behind in the calculation of characteristic equation should be shut. A 'backdoor-free' complete set can be obtained via unitary transformation.

22. With invariance of parity operator we mean $P^2 \equiv 1$ in any orthonormal representations. The prerequisite for $P^2 = 1$ to hold is that the bosonic part $e^{i\pi \sum_k a_k^+ a_k}$ satisfies $e^{i\pi \sum_k a_k^+ a_k} e^{i\pi \sum_k a_k^+ a_k} = 1$.

A complete set is the sufficient condition for the invariance, but not the necessary condition. For instance, the parity operator with regard to the Hamiltonian $H_{sb}$ in eq.(1), $P_{sb} = \sigma_x e^{i\pi \sum_k a_k^+ a_k}$, its bosonic part can be expanded in the Fock states as

$$e^{i\pi \sum_k a_k^+ a_k} = e^{i\pi \sum_k a_k^+ a_k} \sum_{\{n\}=0}^{\{\infty\}} |\{n\}\rangle\langle\{n\}| = \sum_{\{n\}=0}^{\{\infty\}} (-1)^{\{n\}} |\{n\}\rangle\langle\{n\}|.$$ Since

$$e^{i\pi \sum_k a_k^+ a_k} e^{i\pi \sum_k a_k^+ a_k} = \sum_{\{m\}=0}^{\{\infty\}} (-1)^{\{m\}} |\{m\}\rangle\langle\{m\}| \sum_{\{n\}=0}^{\{\infty\}} (-1)^{\{n\}} |\{n\}\rangle\langle\{n\}|$$

$$= \sum_{\{m\},\{n\}=0}^{\{\infty\},\{\infty\}} (-1)^{\{m\}+\{n\}} |\{m\}\rangle\langle\{m\}|\{n\}\rangle\langle\{n\}|$$

$$= \sum_{\{n\}=0}^{\{\infty\}} (-1)^{2\{n\}} |\{n\}\rangle\langle\{n\}| = \sum_{\{n\}=0}^{\{\infty\}} |\{n\}\rangle\langle\{n\}| = 1$$

Thus the completeness guarantees $P_{sb}^2 \equiv 1$. But even in a subspace, the expansion $e^{i\pi \sum_k a_k^+ a_k} = \sum_{\{n\}=0}^{\{N_{tr}\}} (-1)^{\{n\}} |\{n\}\rangle\langle\{n\}|$, since all the non-zero elements are on the diagonal, and of which the absolute value is 1, it always has $e^{i\pi \sum_k a_k^+ a_k} e^{i\pi \sum_k a_k^+ a_k} = \sum_{\{n\}=0}^{\{N_{tr}\}} |\{n\}\rangle\langle\{n\}| = I$, i.e., it remains a unit matrix. Thus $P_{sb}^2 \equiv \sum_{\{m\}=0}^{\{N_{tr}\}} |\{m\}\rangle\langle\{m\}| \sum_{\{n\}=0}^{\{N_{tr}\}} |\{n\}\rangle\langle\{n\}| \sigma_x \sigma_x = I \sigma_x \sigma_x = 1$, any finite expansion in the Fock states can also guarantee the invariance of the parity operator. However, in this case the characteristic equation for the eigenvalues problem is underdeterminate.

The displaced Fock states $|\{n\}\rangle_A$ guarantees that the characteristic equation for the eigenvalues problem of spin-boson model has single unique equation under any truncation, but only in the complete set of this representation the parity operator satisties $P^2 \equiv 1$. This is because the matrix element of the bosonic part of the parity operator in this representation (eq.(18)) is nonzero. For a finite expansion, $P^2 \approx 1$ holds only for small α. Over a critical value of α, $P^2 \approx 1$ is not valid, and parity breaking behavior ensues.

The Fock states $\{|\{n\}\rangle\}$ are independent from the electron-phonon interaction, therefore in finite expansion of Focck states the $P_{sb}^2 \equiv 1$ can still hold, but the displaced Fock states involve electron-phonon interaction, this inevitably, at large dissipation parameter, causes the bosonic part of the parity operator under finite expansion of this representation to have $P^2 < 1$, thus the parity symmetry is broken as artifact of finite expansion.

For representation of the parity operator in some spaces, e.g., the Fock states, the finite expansion could be complete, whereas in other spaces, e.g., the coherent states, the finite expansion could not be complete. In doing numerical calculation, with sufficiently higher orders included, the eigenvalues could be taken as +1 and −1 (in the sense of a real number as in practical computation), the parity operator seems to be complete. However, in Fock state, though the completeness of parity operator can be preserved in finite expansion, the stationary Schrödinger equation for the system in the finite space is an underdetermined equation *(the number of undetermined coefficients is larger than the number of independent equations. In the multiple mode case, the situation is very serious that under finite truncation, the increasing of undetermined coefficients brought by the increasing bosonic modes is much more than the number of independent equations)*, consequently the stationary equation cannot be solved. Therefore, to make the stationary Schrödinger equation to have single unique solution is the prerequite for the strict discussion of eigenvalue problem for Hamiltonian. Theoretically, the stationary Schrödinger equation for spin-boson model can always be made to have single unique solution via unitary transformation. For instance, for any truncation, eq.(4a) or eq.(4b) has single unique solution.